\documentclass[prb,aps,epsf,twocolumn,showpacs,10pt]{revtex4-1}
\usepackage{graphicx,amsfonts,times,bm,amsmath,verbatim,color,array}
\usepackage{hyperref}
\hypersetup{
	colorlinks=true,final=true,
	linkcolor=blue,
	citecolor=blue,
	filecolor=blue,
	urlcolor=blue,
}

\begin{document}
\title{Enhancement of nuclear spin coherence times by driving dynamic nuclear polarization at defect centers in solids}

\author{Girish Sharma$^1$}
\author{Torsten Gaebel$^2$}
\author{Ewa Rej$^2$}
\author{David J. Reilly$^2$}
\author{Sophia E. Economou$^1$}
\author{Edwin Barnes$^1$}

\affiliation{$^1$Department of Physics, Virginia Tech, Blacksburg, VA 24061, U.S.A\\
$^2$ARC Centre of Excellence for Engineered Quantum Systems, School of Physics,
University of Sydney, Sydney, New South Wales 2006, Australia}

\begin{abstract}
The hyperpolarization of nuclear spins can enable powerful imaging and sensing techniques provided the hyperpolarization is sufficiently long-lived. Recent experiments on nanodiamond $^{13}$C nuclear spins demonstrate that relaxation times can be extended by three orders of magnitude by building up dynamic nuclear polarization (DNP) through the driving of electron-nuclear flip-flop processes at defect centers. This finding raises the question of whether the nuclear spin coherence times are also impacted by this hyperpolarization process. Here, we theoretically examine the effect of DNP on the nuclear spin coherence times as a function of the hyperpolarization drive time. We do this by developing a microscopic theory of DNP in a nuclear spin ensemble coupled to microwave-driven defect centers in solids and subject to spin diffusion mediated by internuclear dipolar interactions. We find that, similarly to relaxation times, the nuclear spin coherence times can be increased substantially by a few orders of magnitude depending on the driving time. Our theoretical model and results will be useful for current and future experiments on enhancing nuclear spin coherence times via DNP.
\end{abstract}

\maketitle

\section{Introduction}
{Quantum coherence in many-body systems can give rise to a variety of intriguing phenomena such as interference and entanglement, which do not have any classical analogue. 
This property lies at the heart of quantum mechanics, and is central to all quantum-based technologies~\cite{Benioff:1980,Feynman:1982,DiVincenzo:1995,Giovannetti:2011,Caves:1981,Ladd:2010,Deutsh:1985,Gisin:2002,Budker:2007}.
However in reality, a quantum  system is never perfectly isolated, but is in contact with its surroundings, to which it loses quantum information via a process called quantum decoherence~\cite{Zeh:1970,Schlosshauer:2005,Leggett:1987,Prokofev:2000,Zurek:2003}.} 
The loss of quantum coherence for nuclear spins remains an important problem in diverse fields such as high resolution magnetic resonance imaging and quantum computation. While on one hand decoherence can lead to broadening of the NMR peak linewidth, it also can lead to errors in the processing of quantum information in systems where nuclear spin states serve as qubits or quantum memories~\cite{ Balestro:2014,Jelezko:2004,Neumann:2008,B. E. Kane,Morton:2008, Taminiau:2014,Ziem:2018,Pfender:2017,Jakobi:2017,Awschalom:2018,Pla:2013}.
Decoherence in this case results from the interaction of the nuclear spin with various environmental degrees of freedom, with the most prominent channel being the interaction of the nuclear spin with other neighboring entities such as another nucleus via the dipole-dipole interaction or electron spin centers via the hyperfine interaction~\cite{Pla:2013,Taminiau:2014,Sar:2012,Thiele:2014}. Nuclear spin-flips resulting from these interactions can induce the loss of information stored in spins and lead to decoherence.
{ Electron-nuclear hyperfine interactions also play an important role in the decoherence of electron spins~\cite{Khaetskii:2002,Khaetskii:2003,Yao:2006,DasSarmaWitzel,Pla:2012,Jelezko:2004,Coish:2004,Maze:2008,Malet:2012,Stanwix:2010,Gaebel:2004}.} Therefore it is of fundamental and technological interest to control and manipulate the underlying nuclear spin degrees of freedom in order to extend the coherence time of nuclear spins, and consequently of electron spins. 

In addition to being a source of decoherence, the electron-nuclear hyperfine interaction can also be used to manipulate the state of the nuclear spin ensemble and in particular to create dynamic nuclear polarization (DNP)~\cite{Abragam:1978}. DNP refers to the generation of a significant nuclear spin polarization through dynamic processes such as driving an electronic system with an external field and subsequent spin polarization transfer from electrons to the nuclei. This technique was initially suggested by A. Overhauser~\cite{Overhauser:1953} in 1953. { He proposed that the large Boltzmann polarization of unpaired electrons could be transferred to neighboring nuclei by saturating the corresponding EPR transition, which could result in an enhanced NMR signal by almost two orders of magnitude~\cite{Overhauser:1953,Carver:1953,Carver:1956}.}
DNP is currently a powerful method that is applicable to a wide variety of physical systems and applications~\cite{Hausser:1968,Wind:1985,Foletti:2009,Sharma2017}. In solids DNP can be achieved by driving the system with optical fields or with microwave fields oscillating close to the electron Larmor frequency, and as a result the large spin polarization of the electrons is transferred to nuclei via a process known as the solid effect~\cite{Abragam:1958,Scheuer:2016, vonMorze:2014, Sukheno:1985, Tateishi:2014, Alvarez:2015, Wang:2016, Rej:2015, Waddington:2017, London:2013, King:2010, Chen:2015, HJWang:2013}.
The polarization created in the nuclear spin ensemble via driven DNP processes will gradually relax and decohere due to internuclear dipolar interactions and other environmental factors.

Let us consider a simple two level system which consists of a single spin in the presence of an external magnetic field $\mathbf{B}=(B^x_\perp(t), B^y_\perp(t), B_\parallel)$ in a perfectly isolated environment. $(B^x_\perp(t), B^y_\perp(t)) = B_\perp(\cos(\nu t), \sin(\nu t))$ is a rotating field in the $x$-$y$ plane
with frequency $\nu$, while $B_\parallel$ is held constant. { The spin Hamiltonian is given by $H=-\boldsymbol{\mu}\cdot\mathbf{B}$, where $\boldsymbol{\mu}=-{g}\mu_n\boldsymbol{\sigma}$ is the magnetic moment, and $\boldsymbol{\sigma}$ is the vector of Pauli spin matrices. Explicitly, the Hamiltonian can be written as 
\begin{align}
H=\left( \begin{array}{cc}
\omega/2 & \Omega e^{-i\nu t} \\
 \Omega e^{+i\nu t} & -\omega/2  \\
\end{array} \right),
\end{align}
where $\omega = g \mu_n B_\parallel$, and $\Omega = g\mu_n B_\perp/2$.}
The spin dynamics under this Hamiltonian is governed by the von Neumann
equation for the density matrix $ {\dot {\rho}}=-i[H,\rho]$. {In the frame co-rotating with the magnetic field}, the equations of motion for the components of the density matrix can be written as $ i {\dot {\rho}}_{\uparrow \uparrow}=\Omega (\rho_{\uparrow 
\downarrow}^{*}-\rho_{\uparrow \downarrow})$, 
and $i {\dot {\rho}}_{\uparrow \downarrow}= \delta \rho_{\uparrow \downarrow}+\Omega (\rho_{\downarrow \downarrow}-\rho_{\uparrow\uparrow})$
where $\delta = \omega - \nu$ is the detuning frequency of the applied
transverse rotating field,
$\Omega$ is the corresponding Rabi frequency, and the arrow represents the direction of the spin. The time evolution of this spin system is perfectly unitary since we have assumed the spin is perfectly isolated from its environment. Consequently in this ideal situation, the coherence time is infinite as the phase information is always maintained. However as we discussed above, coupling to various environmental degrees of freedom needs to be accounted for in the Hamiltonian, which leads to loss of spin coherence over time. Due to the complicated and many-particle nature of this problem, which involves several microscopic environmental parameters, a simpler phenomenological approach to this problem is to add decay terms in the equations i.e.  $i{\dot {\rho}}_{\uparrow \uparrow}=\Omega (\rho_{\uparrow 
\downarrow}^{*}-\rho_{\uparrow \downarrow}) -{i}\rho_{\uparrow \uparrow}/{T_{1}}$, and $i {\dot {\rho}}_{\uparrow \downarrow}= \delta \rho_{\uparrow \downarrow}+\Omega (\rho_{\downarrow \downarrow}-\rho_{\uparrow\uparrow})-{i}\rho_{\uparrow
\downarrow}/{T_{2}}$. This phenomenological approach successfully  describes a variety of experiments~\cite{Slitcher:1990, Sham:1999}. {Although the Bloch equations do not work well when the environment cannot be described by classical white noise as in the present work, the parameters $T_1$ and $T_2$ still characterize longitudinal and transverse relaxation respectively in the two level system and can be extracted from experimental measurements.}

It is the transverse dephasing time $T_2$ which characterizes how long a quantum state remains coherent.
In the fields of magnetic resonance imaging and quantum computation, pulse sequence techniques have been developed as a method to reduce spin dephasing and thereby increase the $T_2$ coherence time~\cite{Haeberlen:1976, Vandersypen:2004, Viola:1998, Viola:2004, Chen:2006, Lee:2007, Khodjasteh:2005, Uhrig:2007}. Some well known examples of these include Hahn's spin echo (SE) (single $\pi$ pulse), the Carr-Purcell-Meiboom-Gill (CPMG) pulse sequence (multiple $\pi$ pulses), and periodic dynamical decoupling (PDD). These control techniques extend coherence times by effectively decoupling the system from its environment~\cite{Haeberlen:1976, Vandersypen:2004, Viola:1998, Viola:2004, Chen:2006, Lee:2007, Khodjasteh:2005, Uhrig:2007}.

In semiconductors, it has been shown that electron spin dephasing times ($T_{2e}^*$) can be extended by programming the nuclear spin ensemble as this effectively reduces the number of degrees of freedom and, consequently, the variance in effective fields sampled by an ensemble of experiments~\cite{Reilly:2008, Foletti:2009, Bluhm:2010}. { It has also been shown that polarizing electron spins in ensembles can improve electron spin coherence times ($T_{2e}$)~\cite{Takahashi:2008}.} DNP is also expected to enhance electron spin coherence times ($T_{2e}$)~\cite{Barnes:2012}, although this has yet to be demonstrated in experiment.
Recent experiments have demonstrated an increase in the relaxation time of nanodiamond $^{13}$C spins by dynamically polarizing the nuclear bath via microwave assisted DNP~\cite{Rej:2015}. The nuclear spin relaxation time has been observed to increase with the hyperpolarization time by 3 orders of magnitude. This raises the question of whether the nuclear spin coherence time is similarly impacted by this hyperpolarization process. { Preliminary experimental results on nanodiamond $^{13}$C spins suggest that the hyperpolarization process can in fact enhance the nuclear spin coherence time~\cite{Expt,aps}.} 

In this work we theoretically examine the effect of DNP on nuclear spin coherence and calculate $T_2$ as a function of the driving time. We start with a central spin model of an electron located at a paramagnetic site interacting with the surrounding nuclear bath via the hyperfine interaction. Driving this system at a microwave frequency close to the ESR frequency then induces a large DNP in the surrounding nuclear spin bath, which typically spreads out to a few nanometers around the electron site. Using Liouville's equation, we then calculate the spatial distribution of the nuclear polarization around the electron site as a function of the driving time. The DNP induced effective magnetic field produced by polarization of the nuclear spins in the crystal however also undergoes its own dynamics due to the internuclear dipole-dipole coupling~\cite{Lowe:1967, Paget:1982}. These dynamics lead to local fluctuations in the nuclear spin field. Since the exact quantum mechanical treatment of this mechanism (which is a many-particle dipolar interaction) is prohibitively complicated, in this work we treat the dynamical evolution of nuclear spins caused by such an interaction using a stochastic diffusion model for an effective nuclear spin field~\cite{Filip:2017, Gong:2011}, which remains a valid approximation for the time scales considered in this work. { The use of stochastic models to describe diffusion in dipolar-coupled spin lattice systems has a long history dating back to the seminal works of Anderson and Klauder~\cite{AndersonWeiss,KlauderAnderson}. Here, we extend these approaches to include the diffusion of inhomogeneous DNP originating from a driven electronic spin.} Using an effective Gaussian model for the DNP solution, we then analytically calculate the nuclear spin-spin correlation function, which is crucial for the evaluation of the $T_2$ time. Finally, we calculate the $T_2$ coherence time and study its dependence under dynamical decoupling pulse sequences (like SE, CMPG) as a function of the driving time. We find that the nuclear $T_2$ coherence time substantially increases by a few orders of magnitude depending on the DNP drive time. We also apply our results to calculate the $T_2$ time of $^{13}$C nuclear spins for experimentally relevant parameters. Our theoretical model and results will be useful for current and upcoming experiments on enhancing the coherence time of nuclear spins via DNP. 


This paper is organized as follows: In Sec II we discuss dynamic nuclear polarization in solids and obtain an exact numerical solution for the spatial distribution of nuclear polarization for a microwave driven DNP process enabled by electron-nuclear hyperfine interaction. In Sec III we discuss
the problem of diffusion of the spin polarization into the bulk of the sample within the framework of a stochastic model, and also calculate the two point correlation function. In Sec IV we evaluate the nuclear spin coherence time as a function of the polarization time. We conclude in Sec V.


\section{Dynamic nuclear polarization}
\begin{figure}
\includegraphics[width=\columnwidth]{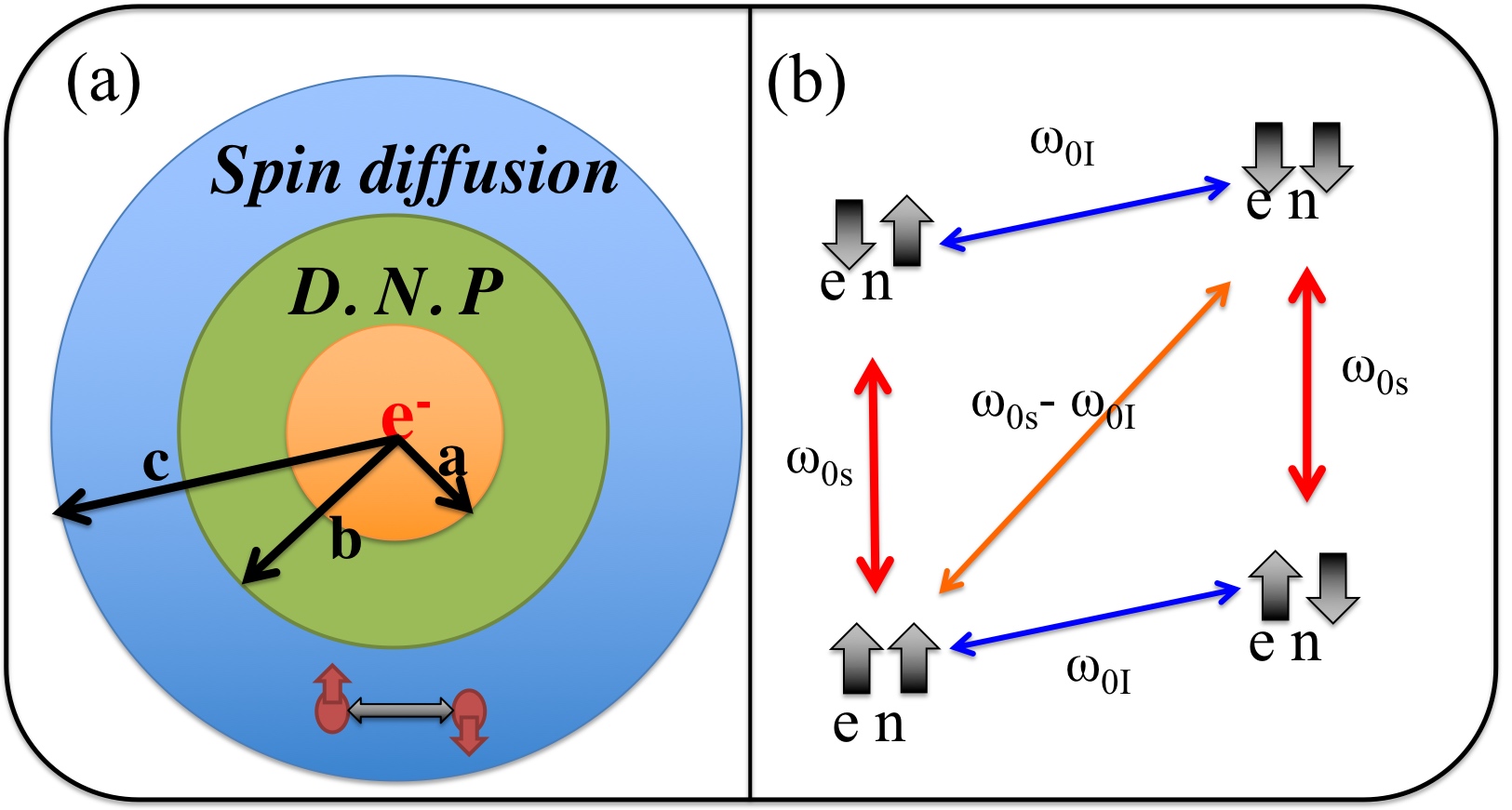}
\caption{(a) A schematic representation of polarization transfer via DNP. The electron center constitutes an ESR line at a frequency $\omega_{0s}$. The nuclear spins in a radius of $r<a$ around the electron center are strongly polarized while the nuclei within $a<r<b$ participate in the DNP process. The spin polarization is transferred further away to $b<r<c$ via nuclear spin diffusion mediated by dipole-dipole interactions between the nuclei. In red is shown a possible nuclear process where there is a spin flip-flop between pairs of nuclei, leading to local fluctuations in the nuclear field. (b) The corresponding energy level diagram. The ESR line is at frequency $\omega_{0s}$ and the NMR line is at frequency $\omega_{0I}$. Driving the system at a microwave frequency $\omega_m=\omega_{0s}-\omega_{0I}$ results in a spin polarization transfer between electron and nuclei (DNP).}
\label{Fig_schematic_dnp}
\end{figure}

Let us begin by first considering the case of a single unpaired electron (at a paramagnetic site) in a solid located at the origin which is coupled to the surrounding nuclear bath in the presence of a magnetic field. The total Hamiltonian for this system is given by
\begin{align}
H = H_e + H_n +H_{en} + H_{nn},
\end{align}
where $H_e = \omega_{0s} S_z$ and $H_n=- \sum\limits_i\omega_{0I}I^i_z $ are the Zeeman energies of the electron and the nuclei respectively in the presence of an external magnetic field. The summation $i$ is over all the nuclei, which are assumed to be of the same species for simplicity. $H_{en}$ is the hyperfine interaction between electron and nuclei, and $H_{nn}$ is the dipole-dipole interaction between the nuclei. The hyperfine coupling between the electron and nuclei comprises an isotropic contact hyperfine interaction $\sim a_{iso} S\cdot I$ (which requires a non-zero overlap between the electron and nuclear wave functions and is therefore localized near the electron site), and the anisotropic interaction given by the dipolar coupling between the electron and nuclear magnetic moments. For our purposes we will only consider the dipolar coupling, retaining parts which are significant under the high magnetic field approximation~\cite{Slitcher:1990}: 
{
\begin{align}
H_{en}= \sum_i A_1^i S_z I_z^i+A_2^i S_z I_x^i,
\label{Eq:Hen_1}
\end{align} 
where $A_1^i$ and $A_2^i$ are the secular and non-secular parts of the interaction respectively~\cite{Slitcher:1990}
\begin{align}
A_1^i &= T (3\cos^2\theta_i-1),\\
A_2^i &= 3T \sin\theta_i\cos\theta_i,
\label{Eq_A1_A2}
\end{align}}
where $T = {(\mu_0/4\pi r^3)}\gamma_e\gamma_N \hbar^2$ is the bare hyperfine strength, $r$ is the distance between the electron center and nuclear spin, $\theta_i$ is the polar angle, $\gamma_e$ and $\gamma_N$ are the electron and nuclear gyromagnetic ratios. 
{ As depicted in Fig.~\ref{Fig_schematic_dnp}(a), the nuclear spins surrounding the electronic center can be divided into three regions according to their role in the DNP process. The nuclei very close to the electronic spin (i.e., those within radius $r<a$, where $a$ is typically 0.1 nm to 0.3 nm~\cite{Wolfe:1973}) undergo rapid electron-nuclear spin flip-flops (second term in Eq.~\ref{Eq:Hen_1}) and become polarized. However, they also experience a strong Knight shift (first term in Eq.~\ref{Eq:Hen_1}) that changes their effective Zeeman energies. This prevents these nuclei from participating in nuclear spin diffusion because the diffusion process is driven by energy-conserving nuclear spin dipolar flip-flop processes originating from the internuclear dipole-dipole interaction $H_{nn}$. This is known as the frozen-core or diffusion-barrier effect~\cite{Wolfe:1978,Reynhardt:1998,Ramanathan:2008}. The nuclear dipolar flip-flops are suppressed by the energy mismatch between the nuclei close to the defect center and those in the bulk. Thus, it is the nuclei which are further away from the center ($a<r<b$) but still hyperfine-coupled to the electronic spin that are responsible for the diffusion process. The distance $b$, which depends on the transition probabilities of the hyperpolarization mechanisms and relaxation, can typically extend up to a few nanometers~\cite{Reynhardt:1998,Hoch:1988}. The nuclei in this range are still close enough to the center to undergo electron-nuclear hyperfine flip-flops and become polarized, but they are far enough away that their Knight shifts are small and do not prevent dipolar flip-flops with nuclei in the bulk. In Fig.~\ref{Fig_schematic_dnp}, we represent the bulk nuclei as those lying in the radial span $b<r<c$. 

In this section, we focus on quantitatively understanding how the nuclei in the intermediate region $a<r<b$ become polarized. We are particularly interested in the case where this DNP process is accelerated by external microwave driving. Note that the rate and amount of DNP will vary across the intermediate region since the hyperfine interaction $H_{en}$ scales like $r^{-3}$ and has an angular dependence. Here, we determine the distribution of nuclear spin polarization across this region as a function of the driving time. In the next section, we then study how this nuclear polarization gets transferred to the bulk via spin diffusion.}

\begin{figure}
\includegraphics[width=\columnwidth]{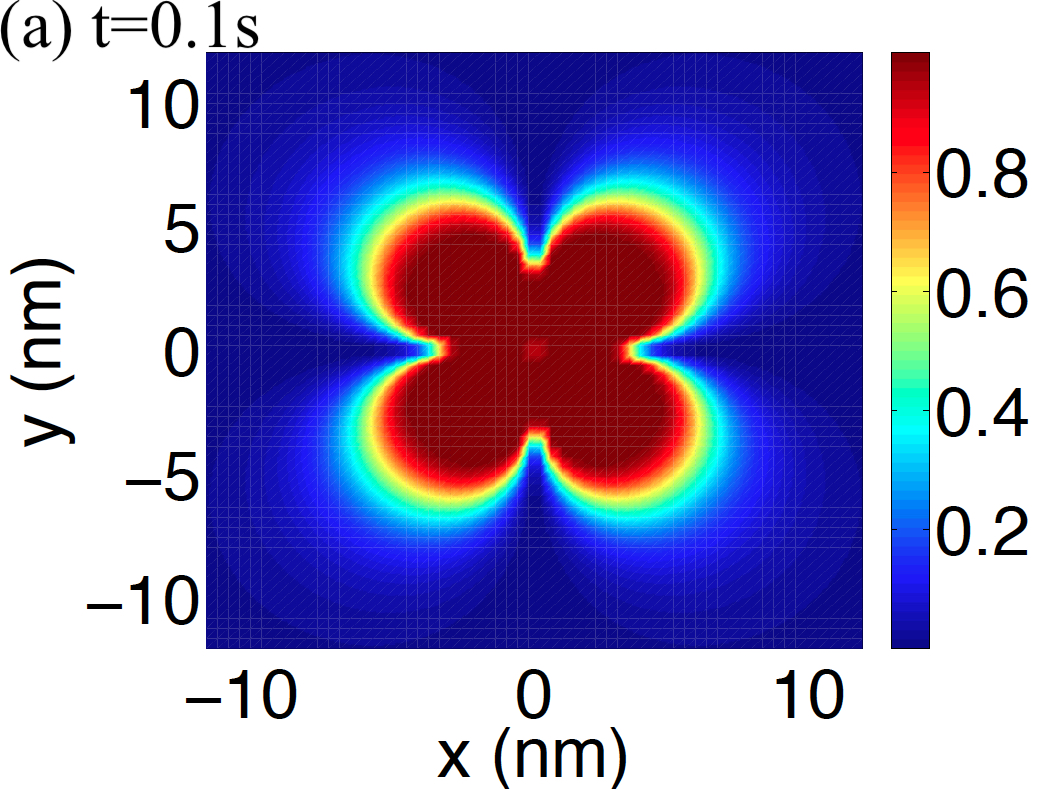}
\includegraphics[width=\columnwidth]{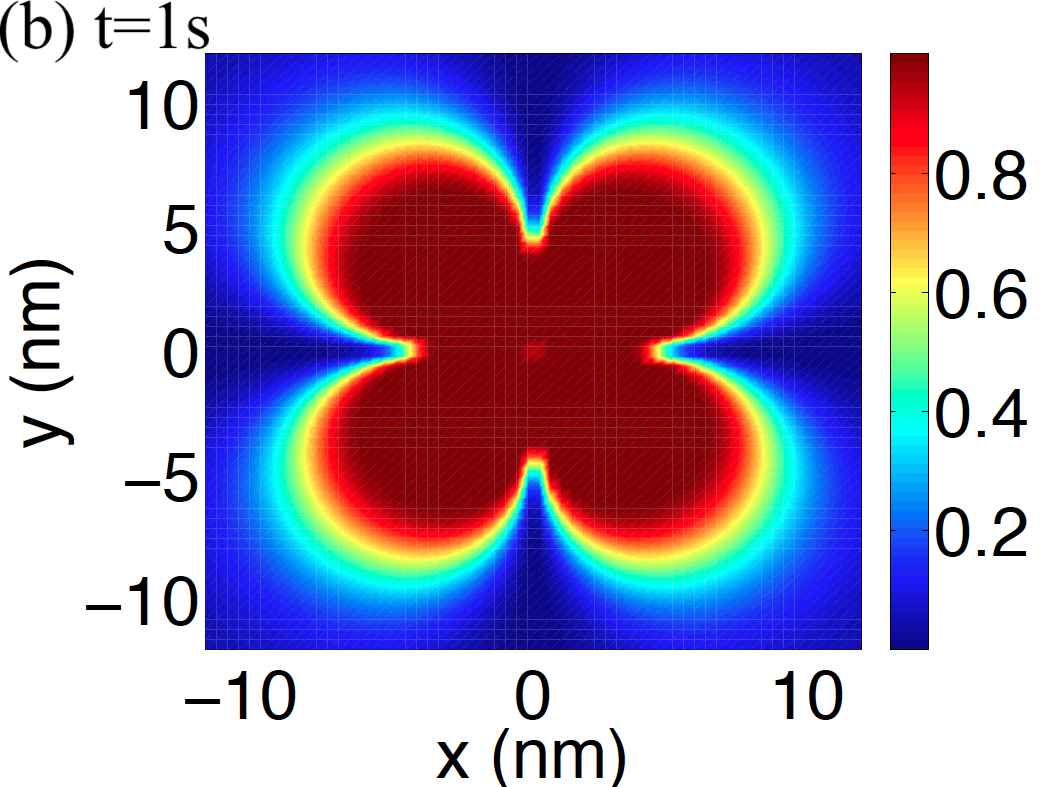}
\caption{Spatial profile $\langle P_z(r,\theta,t)\rangle$ of the microwave driven nuclear polarization around the electron center at the origin as obtained from the solution of the Liouville-von Neumann equation at various times. \textit{(a)} At time $t=0.1$s. \textit{(b)} At time $t=1$s. We use the parameters $\omega_{0s}=10\mu$eV, $B_e=0.1\mu$eV, $\omega_{0I}=0.0004\omega_{0s}$. The relaxation parameters for the Lindblad operator used were $\gamma_1=1\mu$s, and $\gamma_2=1$s. Note that this so far does not account for nuclear diffusion (which is the subject of Sec III), but rather provides an initial condition to that problem.}
\label{Fig_dnp_spatial_1}
\end{figure}

We first consider the case of an electronic center coupled to a single nuclear spin. Diagonalizing the Hamiltonian $H_0=H_e+H_n+H_{en}$ 
\begin{align}
\tilde{H_0} &= \nonumber U H_0 U^{\dagger}\\ 
&=\omega_{0s} S_z - \tilde{\omega}_{0I}I_z + A'S_zI_z ,
\end{align}
where 
\begin{align}
\tilde{\omega}_{0I}  &= \frac{\omega_{0I}}{2}\left(\cos \eta_- + \cos \eta_+ \right) - \frac{A_1}{4} \left(\cos \eta_- - \cos \eta_+ \right) \nonumber\\&- \frac{A_2}{4} (\sin \eta_- - \sin \eta_+),
\end{align}
\begin{align}
A'  &= -{\omega_{0I}}\left(\cos \eta_- - \cos \eta_+ \right) + \frac{A_1}{2} \left(\cos \eta_- + \cos \eta_+ \right) \nonumber \\&+ \frac{A_2}{2} (\sin \eta_- + \sin \eta_+),
\end{align}
\begin{align}
\tan \eta_{\mp} = \frac{A_2}{A_1 \mp 2\omega_{0I}}, 
\end{align}
and $U = \exp[i (\eta_- (I/2 + S_z) I_y + \eta_+ (I/2 - S_z)I_y)]$ is the unitary transformation.
To study DNP in this system, we now introduce the microwave field $H_M = B_e S_x \cos(\omega_m t) $, which can drive the nuclear spin flip transitions for particular values of the drive frequency $\omega_m$. 
{ Here, we neglect the direct effect of this oscillating field on the nuclear spins because their Zeeman splitting is on the order of MHz and thus far detuned from the drive. However, the microwave drive can still indirectly cause nuclear spin flips due to hybridization of the electronic and nuclear spin states under the hyperfine flip-flop interaction. These hybridized states are the eigenstates of $H_0$. Although these states are hybridized, we can still label them in terms of separate spin quantum numbers for the electron and nucleus since the hybridization is weak owing to the smallness of the hyperfine coupling constant. Thus we can denote them as $|{s_es_n}\rangle$, where $s_e$ and $s_n$ each assume the values $\uparrow$, $\downarrow$ corresponding to the spin. We can see how these states are affected by the drive by transforming $H_M$ into the eigenstate basis~\cite{Hu:2011} :}
\begin{align}
\tilde{H}_M &= \nonumber U H_M U^{\dagger}\\   
&=B_e \cos(\omega_m t) (S_x \cos \delta - \frac{1}{2}{(S_+I_- + S_-I_+)} \sin\delta \nonumber \\&+ \frac{1}{2}{(S_+I_+ + S_-I_-)} \sin\delta),
\end{align}
where $2\delta = \eta_--\eta_+$. 
{ The $S_{\pm}I_{\pm}$ terms in the above Hamiltonian are the spin-flip terms that couple different eigenstates of $H_0$ and enable the generation of DNP. In particular, they drive transitions between the states $|{\uparrow\uparrow}\rangle$ and $|{\downarrow\downarrow}\rangle$ and between $|{\uparrow\downarrow}\rangle$ and $|{\downarrow\uparrow}\rangle$.} When $A_2=0$, these spin flip terms in $\tilde{H}_M$ vanish, indicating the absence of any microwave driven nuclear polarization. Note that from Eq.~\ref{Eq_A1_A2}, $A_2$ is generically non-zero except when $\theta=n\pi/2$, where $n=\{0,1,2,3\}$. When $A_2\neq 0$, either the $S_+I_- + S_-I_+$ or the $S_+I_+ + S_-I_-$ terms facilitate the transfer of nuclear polarization. In the interaction frame of $\tilde{H}_0$, one can derive an effective Hamiltonian $H_{\text{eff}}$ when the microwave frequency is tuned to drive either of the nuclear spin flip transitions. { In the remainder of the paper, we focus on the latter choice, and the corresponding energy levels and drive frequency are depicted in Fig.~\ref{Fig_dnp_spatial_1}(b).} Specifically, when $\omega_m=\omega_{0s}-\omega_{0I}$ the double spin-flip term $S_+I_+ + S_-I_-$ in $\tilde{H}_M$ is selected and $H_{\text{eff}}$ becomes
\begin{align}
H_{\text{eff}} = \frac{B_e}{4} \sin\delta (S_+I_+ + S_-I_-).
\label{Eq_H_eff}
\end{align}
In the interaction frame of reference, microwave driven nuclear dynamics can then be described by the  Liouville-von Neumann equation $\dot{\rho}=-i[H_{\text{eff}},\rho] + L[\rho]$, where we can solve for the evolution of the density matrix $\rho$ corresponding to the 4-component electron-nuclear spin system. The Lindblad operator $L[\rho]$ accounts for relaxation processes (like decay of the electron spin to the ground state), which gives a non-unitary evolution of the quantum system and consequently a steady state solution for $\rho(t)$ when $t\rightarrow\infty$. Specifically $L[\rho]$, which is the relaxation superoperator term, can be expressed as
\begin{align}
L[\rho] = \sum\limits_{k} \left(L_{k}\rho L_{k}^\dagger-\frac{1}{2}[L_{k}^\dagger L_{k}\rho + \rho L_{k}^\dagger L_{k}] \right),
\label{Eq_Lindblad_op}
\end{align}
where $L_{k}$ are the Lindblad operators. The index $k$ runs from 1 to 4, with the non-trivial elements of the Lindblad operators being $\langle \uparrow\uparrow|L_1|\downarrow\uparrow\rangle=\sqrt{\gamma_2}$, $\langle \uparrow\downarrow|L_2|\downarrow\downarrow\rangle=\sqrt{\gamma_2}$, $\langle \uparrow\uparrow|L_3|\downarrow\downarrow\rangle=\sqrt{\gamma_1}$, $\langle \uparrow\downarrow|L_4|\downarrow\uparrow\rangle=\sqrt{\gamma_1}$, where the first (second) arrow indicates the direction of the electron (nuclear) spin. $L_1$ and $L_2$ describe processes conserving nuclear spin, while $L_3$ and $L_4$ describe (much slower) processes involving flipping of the nuclear spin as well. The Liouville-von Neumann  equation can be solved analytically as  $ \tilde{\rho}(t) = \mathcal{S}(t)\tilde{\rho}(0)$, where $\tilde{\rho}(t)$ is the density matrix $\rho(t)$ written  as a single column vector, and the matrix $\mathcal{S}(t)$ is $ \mathcal{S}(t) = e^{(\mathcal{H+G})t}$, where $ \mathcal{H} = i(H_{\text{eff}}\otimes I - I\otimes H_{\text{eff}})$, $ \mathcal{G} = \sum\limits_{m}\left[L_m\otimes L_m - \frac{1}{2}I\otimes L_m^\dagger L_m - \frac{1}{2}L_m^\dagger L_m \otimes I\right]$. Thus for an arbitrary initial condition $\rho(0)$, the density matrix $\rho(t)$ at a later time can be evaluated exactly.

\begin{figure}
\includegraphics[width=\columnwidth]{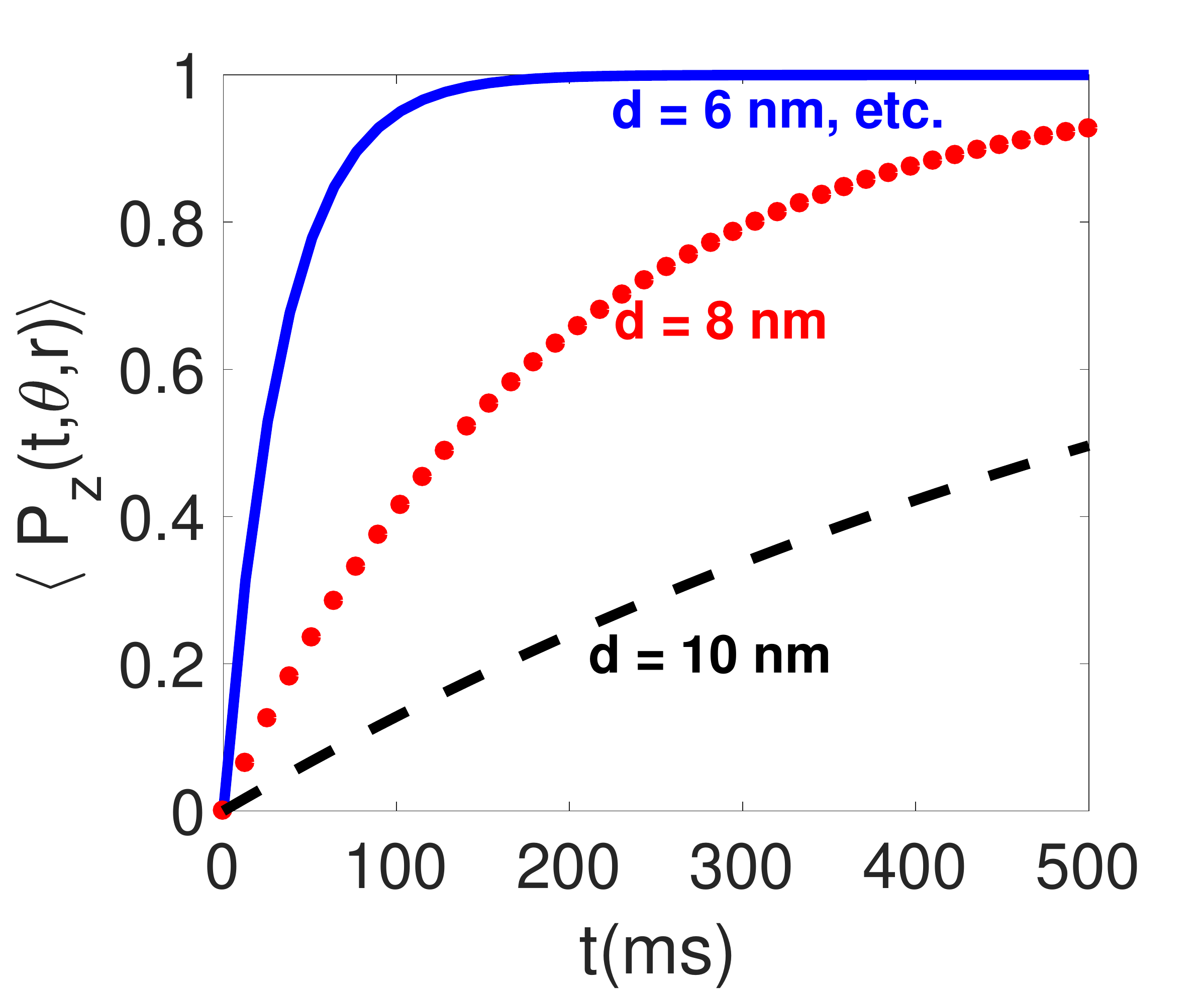}
\caption{Time evolution of DNP around the electron center at the origin as obtained from the solution of the Liouville-von Neumann equation at various distances from the origin. We use the parameters $\omega_{0s}=10\mu$eV, $B_e=0.1\mu$eV, $\omega_{0I}=0.0004\omega_{0s}$, $\theta=\pi/4$. The relaxation parameters for the Lindblad operator used were $\gamma_2=1\mu $s, and $\gamma_1=1$s. Also note that this solution so far does not account for nuclear spectral diffusion (which is the subject of Sec III), but rather serves as its initial condition.}
\label{Fig_dnp_time_1}
\end{figure}

{The spatial profile (which arises from the spatial dependence of the hyperfine constants $A_1$ and $A_2$) of the induced nuclear polarization around the electron center can  be studied as a function of the microwave driving time by evaluating the mean value of nuclear polarization $\langle P_z(t,r,\theta)\rangle=\text{Tr} (P_z\rho(t,r,\theta))$ where the operator $P_z$ is the nuclear spin operator $I_z$ written in the interaction representation. Explicitly, $P_z = e^{i\tilde{H_0} t} U I_z U^\dagger e^{-i\tilde{H_0}t}$.
In Fig.~\ref{Fig_dnp_spatial_1} we plot the spatial profile of the microwave driven nuclear polarization around the electron center at the origin as obtained from the solution of the Liouville-von Neumann equation at two different times. Clearly the spread of DNP with increasing drive times is evident. 
It is evident from the figure that the buildup of DNP is not isotropic. The origin of this pattern can be traced back to the anisotropy of the non-secular part ($A_2$) of the hyperfine interaction (see Eq.~\ref{Eq_A1_A2}). We noted earlier that $A_2$ is responsible for DNP, and its $\theta$ dependence precisely gives rise to the non-uniform pattern in Fig.~\ref{Fig_dnp_spatial_1}.}
Even though the spatial profile around the electron center is not isotropic, the schematic representation presented in Fig.~\ref{Fig_schematic_dnp} provides us with a good approximation to the problem. This DNP solution will be used as an initial condition for the problem of nuclear spin diffusion (which is the subject of Sec III). 
Fig.~\ref{Fig_dnp_time_1} shows the time evolution of DNP at various distances from the origin as obtained from the solution of the Liouville equation.
\section{Nuclear spin diffusion}
We have so far focused on the problem of DNP in solids, which produces a large spin polarization of the nuclei   around the paramagnetic electron center. In this section we will discuss the problem of diffusion of the spin polarization into the bulk of the sample. The large nuclear Overhauser field produced via DNP in the vicinity of this electron defect site dynamically evolves through mutual dipole-dipole coupling of the nuclei. This also causes temporal fluctuations of the nuclear spin field due to processes such as a pair of spin flips.
Since the exact quantum mechanical treatment of this many-particle dipolar interaction is quite complicated, we will describe the dynamical evolution of the nuclear spin field with the following stochastic diffusion model~\cite{Filip:2017}
\begin{align}
\frac{\partial I(\mathbf{x},t)}{\partial t} = D \frac{\partial^2}{\partial \mathbf{x}^2} I (\mathbf{x},t) + \zeta(\mathbf{x},t).
\label{Eq_Diff_1}
\end{align}
In the above equation $I(\mathbf{x},t)$ is the nuclear field, which on a coarse-grained scale encompasses several atomic sites, $D$ is the isotropic diffusion constant, and $\zeta(\mathbf{x},t)$ is an effective stochastic field which models the randomness associated with the nuclear spin flips. { The validity and the derivation of the above model has been a subject of many earlier works~\cite{Lowe:1967,Gong:2011,Deng2005}, but for the sake of completeness we will briefly sketch the derivation of the above diffusion model. 

The Hamiltonian for the mutual dipole-dipole coupling between the nuclei can be written as a sum of the Ising and the flip-flop terms:
\begin{align}
H_{nn} = -\sum_{i\neq j} (B_{ij} I_+^i I_-^j - 2B_{ij} I_z^i I_z^j),
\end{align}
where $B_{ij}$ is the coupling between two nuclear spins $I^i$ and $I^j$, and the summation is over all pairs of nuclear spins. 
We are interested in the evolution of the expectation value for the $z$-component of the nuclear spin at site $k$, which will evolve according to~\cite{Lowe:1967} 
\begin{align}
\frac{\partial}{\partial t} \langle I_z^k\rangle = -\frac{1}{i\hbar} \hbox{Tr}\{\rho(t) [H'_{nn}, I_z^k]\},
\end{align}
where $H'_{nn}=-\sum_{i\neq j} B_{ij} I_+^i I_-^j $ is the nuclear flip-flop interaction, and $\rho(t)$ is the nuclear spin density matrix. This can be rewritten as~\cite{Gong:2011, Lowe:1967, Deng2005} 
\begin{align}
\frac{\partial}{\partial t} \langle I_z^k\rangle = \sum_{i\neq k} W_{ik} ( \langle I_z^i(t)\rangle- \langle I_z^k(t)\rangle),
\label{Eq:dIzbydt_1}
\end{align}
where
\begin{align}
W_{ik} = \int_{0}^{t}{\hbox{Tr}\{ [H'_{nn}(t), I_z^k] [H'_{nn}(t-t'), I_z^i]\}} dt'
\end{align}
has a physical meaning as the flip-flop rate between nuclear spins at sites $i$ and $k$. The parameter $W_{ik}$ can be approximately analytically calculated when the upper integration limit is taken to be large~\cite{Deng2005}. We can then Taylor expand $\langle I_{z}(t)\rangle $ for
site $i$ around site $k$:
\begin{eqnarray}
\langle I^{i}_{z}(t)\rangle &\approx &\langle I^{k}_{z}(t)\rangle +\frac{\partial \langle I^{k}_{z}(t)\rangle }{\partial x^{\alpha }}(x_{k}^{\alpha}-x_{i}^{\alpha })  \nonumber \\
&+&\frac{1}{2}\frac{\partial ^{2}\langle I^{k}_{z}(t)\rangle }{\partial x^{\alpha }\partial x^{\beta }}(x_{k}^{\alpha }-x_{i}^{\alpha
})(x_{k}^{\beta}-x_{i}^{\beta })+\cdots,
\end{eqnarray}
where Einstein’s summation convention is implied for the spatial indices $\alpha$ and $\beta$. Substituting the above equation in Eq.~\ref{Eq:dIzbydt_1}, and identifying $D = D^{\alpha\alpha}=\sum_{i} W_{ik} (x_k^\alpha - x_i^\alpha) ^2/2$, one can obtain the first term on the R.H.S of Eq.~\ref{Eq_Diff_1}.  Note that the linear in $x$ term vanishes.
The connection between the diffusion constant $D$ and the dipole-dipole interaction is now transparent. Note that in Eq.~\ref{Eq_Diff_1}, $I_z(t)$ is simply written as $I(t)$, for ease of notation.} 

{
Our main focus is on natural diamond, which has a 1.1\% concentration of nuclear spins, although our analysis can be applied to the case of enriched diamond as well. A theoretical estimate for the diffusion constant was provided in Refs. ~\onlinecite{Hoch:1988,VanVleck1948,AbragamBook1961}. We summarize the argument here. We start by assuming that the $^{13}$C spins are distributed in a random way in the lattice. A typical distance between two spins may be obtained using the Poisson distribution by putting the probability, $\exp(-(4\pi n_C r^3 /3))$, of finding no other spins within a distance $r$ of a spin placed at the origin, equal to 1/2 ~\cite{Hoch:1988}. Here $n_C$ is the concentration of $^{13}$C. This gives $r =0.55n_C^{-1/3}$, which for natural diamond yields $r =4.42$ \AA. The $^{13}$C spins form a dilute magnetic system for which it is possible to calculate the second moment of the frequency distribution using the Van Vleck expression~\cite{Hoch:1988,VanVleck1948,AbragamBook1961}
\begin{align}
M_2 = \frac{3}{4} \gamma_N^4 \hbar^2 I (I+1) f \sum_k \frac{(1-3 \cos^2 \theta_{jk})^2}{r^6_{jk}},
\end{align}
where $\gamma_N$ is the $^{13}$C gyromagnetic ratio, $I =1/2$, and $f$ is the fraction of lattice sites occupied by magnetic spins (which is 0.011 for natural diamond), and $\theta_{jk}$ is the angle the inter-nuclear vector $r_{jk}$ makes with the magnetic field. For the diamond lattice, this yields $M_2 \sim 1.43 \times 10^6 $s$^{-2}$ (performing the summation for a single crystal over all neighbors and with magnetic field parallel to the [014] direction~\cite{Hoch:1988}). The diffusion constant is given by $D = \frac{1}{30}\sqrt{M_2} r^2$, giving $D \approx 7.8$ nm$^2$/s as a theoretical estimate. The diffusion constant can be obtained experimentally by substituting $M_2=\pi/(2T_2^2)$ in the above formula for the diffusion constant to obtain $D = \tfrac{1}{30}\sqrt{\tfrac{\pi}{2}}r^2/T_2$, where $r$ is again the inter-nuclear separation, and $T_2$ is the measured coherence time (see e.g., Ref. ~\onlinecite{Lowe:1967}). Using the above estimate for $r$ and taking the coherence time to be $\approx 1$ ms~\cite{Expt,aps} yields $D \approx 8.1$nm$^2/$s. The experimental and theoretical estimates agree reasonably well. We employ these estimates to evaluate the coherence time in the next section.

The scale on which $I(\mathbf{x},t)$ varies is on the order of the diffusion length, which can be estimated from the formula $L =\sqrt{2DT_1}$. Using our estimate of the diffusion constant, $D\approx 8.1$ nm$^2$/s, and our previously measured result for $T_1$ (1 hour---see Ref. [57]), we find $L \approx 240$ nm. The fact that this scale is orders of magnitude larger than the inter-nuclear separation ($r\approx 5\hbox{\AA}$ ) justifies the use of a coarse-grained approximation for $I(\mathbf{x},t)$.
}

{ Following Ref.~\onlinecite{Filip:2017}, we treat the field $\zeta(\vec x,t)$ as behaving like white noise, which has the following correlation functions
	\begin{align}
	\langle \zeta(\mathbf{x},t) \rangle &=0,\\
	\langle \zeta(\mathbf{x},t) \zeta(\mathbf{y},s) \rangle & =\Gamma_0 \delta(\mathbf{x}-\mathbf{y})\delta(t-s).
	\label{Eq_Noise_1}
	\end{align}
	White noise is a reasonable assumption since the dipolar flip-flops conserve energy, and the occurrence of one flip-flop event should not alter the probability of a later event.}
In the above equations, averaging is done over all possible noise realizations. The average noise is zero, and choosing $\Gamma_0= -\eta D \partial^2/\partial \mathbf{x}^2$ leads to the conservation of the order parameter~\cite{Hohenburg:1977} (which is the total nuclear spin polarization in the present case), in any possible noise realization. The parameter $\eta$ determines the noise strength. It is convenient to switch to Fourier space, where Eq.~\ref{Eq_Diff_1} becomes
\begin{align}
\frac{\partial}{\partial t} I(\mathbf{q},t) = -D q^2 I(\mathbf{q},t) + \zeta(\mathbf{q},t).
\label{Eq_Diff_2}
\end{align}
The above equation has the following general solution
\begin{align}
I(\mathbf{q},t) = I(\mathbf{q},0)e^{-Dq^2t}+\int\limits_0^t {ds e^{-Dq^2(t-s)}}\zeta(\mathbf{q},s).
\end{align} 
Fourier transforming Eq.~\ref{Eq_Noise_1} we find that the noise correlations in Fourier space take the following form
\begin{align}
\langle \zeta(\mathbf{q},t) \zeta(\mathbf{k},s)\rangle = (2\pi)^3 \eta D q^2 \delta(\mathbf{q}+\mathbf{k})\delta(t-s).
\label{Eq_Noise_2}
\end{align}
Note that the above correlation function vanishes for the zeroth ($\mathbf{q}=0$) Fourier mode, implying a strict conservation of the total nuclear spin polarization. The spin-spin correlation function then becomes
\begin{align}
\langle I(\mathbf{q},t)I(\mathbf{k},s) \rangle &= e^{-Dq^2(t+s)} \langle I(\mathbf{q},0)I(\mathbf{k},0) \rangle \nonumber \\&- \frac{\eta}{2} (2\pi)^3  \delta(\mathbf{q}+\mathbf{k}) (e^{-Dq^2(t+s)} - e^{-Dq^2|t-s|}),
\label{Eq_Iq_Ik}
\end{align}
which can be evaluated once the initial correlation function $\langle I(\mathbf{q},0)I(\mathbf{k},0) \rangle$ at time $t=0$ is known. In the absence of any DNP or any other dynamic processes, this function can be drawn from a stationary equilibrium distribution, however in the present case this initial condition is determined by the solution of DNP (which has been treated in detail in Sec II). The exact functional form $I(\mathbf{x},0)$ of the spatial distribution of DNP (see Fig.~\ref{Fig_dnp_spatial_1}) is not a simple mathematical function with a precise closed form. Hence to make our model analytically tractable for subsequent analysis, we will assume the following simplified form for $I(\mathbf{x},0)$ centered around the defect site (at the origin):
\begin{align}
I(\mathbf{x},0) = I_0 \exp(-\alpha^2 ({x}^2+ y^2 + z^2)).
\label{Eq_I_x0_1}
\end{align}
{ At this point, a few comments are in order. First, note that we are neglecting the role of spin diffusion during the DNP process for the sake of tractability. Including this effect should lead to an enhancement of DNP in the bulk, suggesting that the results we obtain below for the buildup of DNP and its influence on $T_2$ times of bulk nuclear spins may be conservative. However, preliminary experimental results for $T_2$ versus driving time $T$ suggest that this is not a big effect~\cite{Expt,aps} . We also note that we are neglecting nuclear spin relaxation. This is reasonable since our focus is on diffusion and decoherence, both of which occur on timescales much shorter (on the order of seconds) than the relaxation time, which is on the order of minutes or hours~\cite{Rej:2015}.} 

\begin{figure}
\includegraphics[width=\columnwidth]{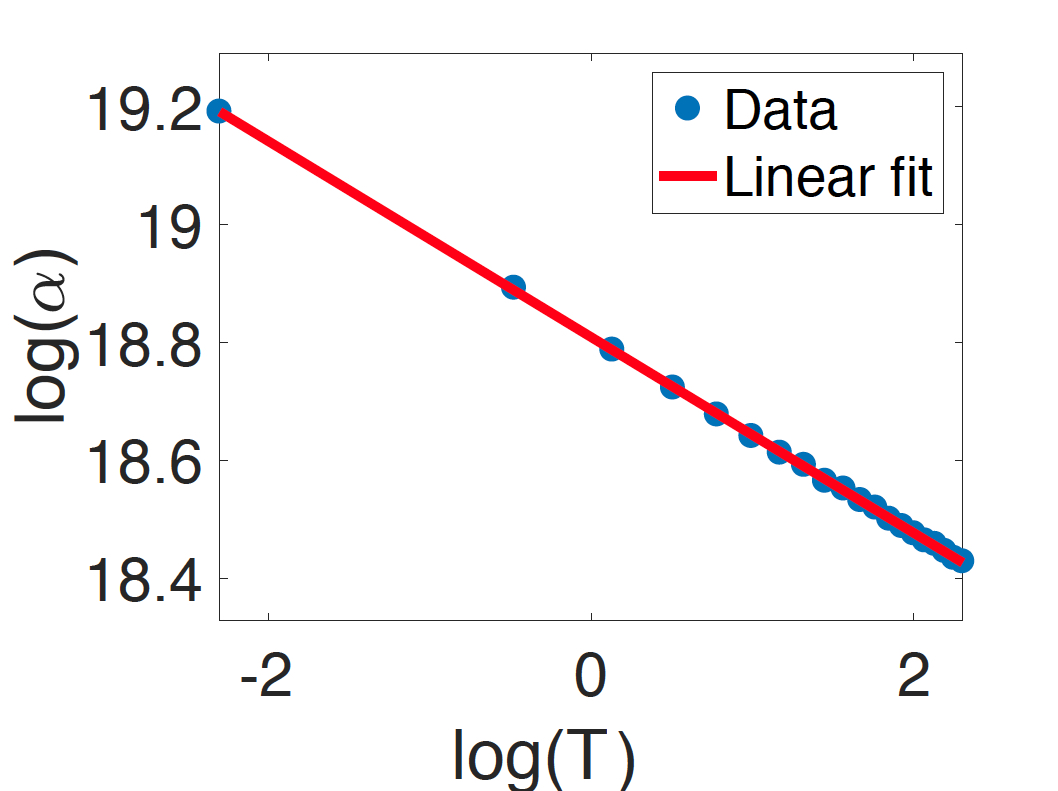}
\caption{Logarithm (natural) of the numerically evaluated $\alpha$ (in the units m$^{-1}$) as a function of the logarithm (natural) of the driving time $T$ (in the units s) using the Gaussian approximation along a particular direction ($\theta=\pi/4$). The plot shows the data (in blue dots) for $\log(\alpha)$ as a function of $\log(T)$. The data shows a linear behavior shown in the red line (at least for the timescales we are concerned with), and thus can be extrapolated to obtain $\alpha$ for arbitrary driving times. The parameters chosen are the same as in Fig.~\ref{Fig_dnp_spatial_1}. We find that for those parameters, $\log(\alpha(T))=a_1 \log{T} +a_2$, with $a_1=-0.16632$, and $a_2=18.8096$. }
\label{Fig_fit_1}
\end{figure}

In Eq.~\ref{Eq_I_x0_1}, the parameter $\alpha$ has a crucial dependence on the DNP  driving time. Specifically, for longer driving times, the spread of the nuclear polarization around the origin should increase, and therefore $\alpha$ should decrease. For a fixed $I_0$ the total nuclear spin polarization ($z-$component) in the sample is $I_0 \pi^{3/2}\alpha^{-3}$, which again highlights the fact that the nuclear polarization increases with driving time. Since we have now approximated the DNP distribution to be of Gaussian form involving the parameter $\alpha$, we must numerically determine $\alpha$ as a function of driving time $T$ using our exact DNP solution from Sec. II. We do this by calculating the full width at half maximum (FWHM) along a chosen direction  from our actual DNP solution (see Fig.~\ref{Fig_dnp_spatial_1}), and relating the calculated FWHM to the standard deviation of our Gaussian approximation, i.e., FWHM $= 2\sqrt{ln2}/\alpha$. Fig.~\ref{Fig_fit_1} shows the plot of the numerically evaluated $\alpha$ as a function of driving time ($T$) using the Gaussian approximation.  Since the plot of $\log(\alpha)$ vs. $\log(T)$ has a linear fit, one can easily obtain $\alpha$ for arbitrary driving times. 

\begin{figure}
	\includegraphics[width=\columnwidth]{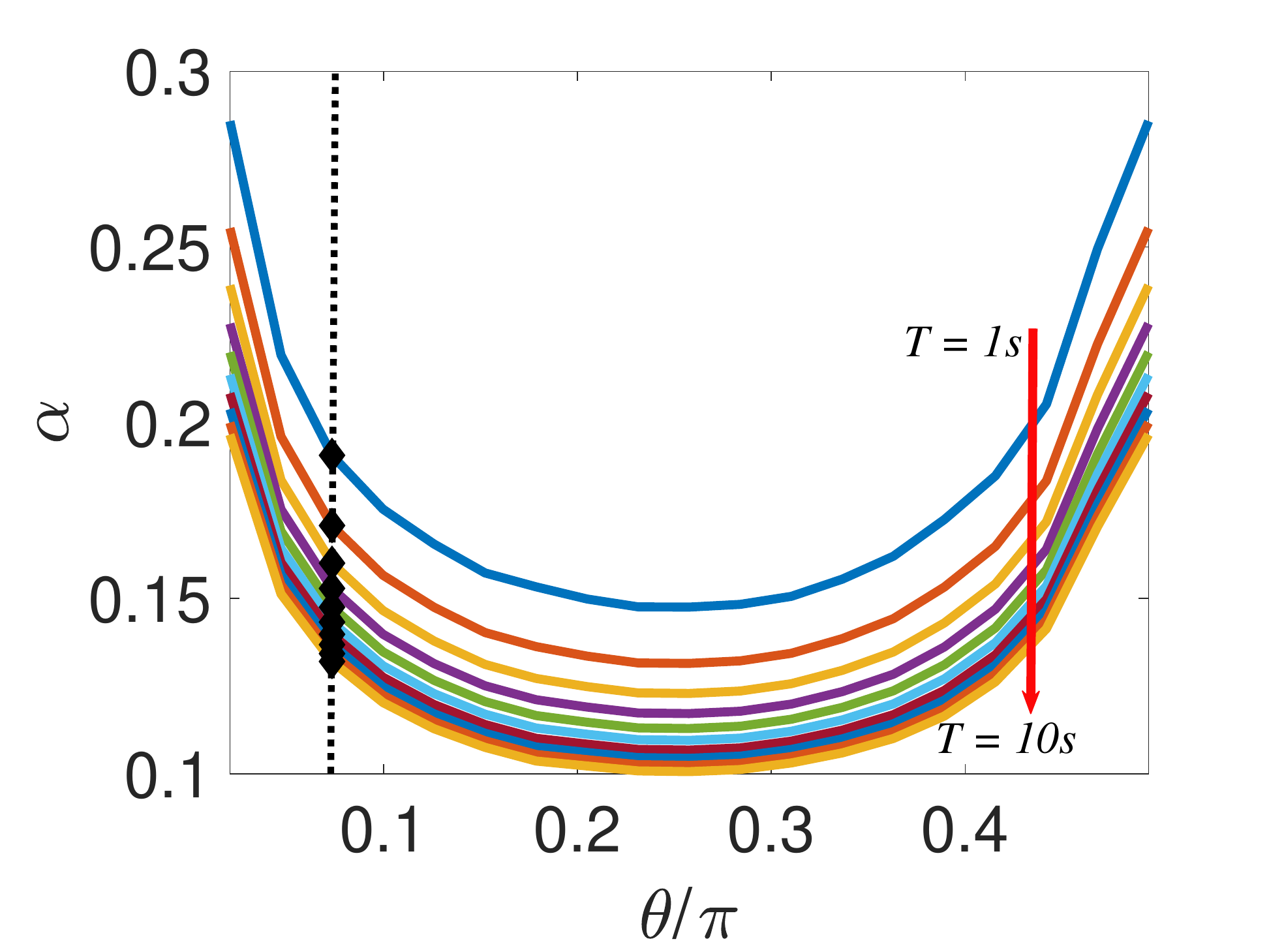}
	\caption{Numerically evaluated $\alpha$ (in the units nm$^{-1}$) as a function of $\theta$ for ten different driving times $T$. The black diamond on each curve indicates the average value $\langle\alpha\rangle_{\theta}$ (obtained by averaging over all $\theta$ values for a particular $T$). The dotted line indicates that the average value for each driving time $T$ collapses on to the value at a particular $\theta$.
		We find that $\langle \alpha\rangle_{\theta}/\alpha_{\theta = \pi/4} \approx 1.24$. We also note that the value of $\alpha$ between $\theta=\pi/4$ and $\theta\approx 0$ differs by just a factor of two. }
	\label{Fig_alpha_2}
\end{figure}

{ Before we proceed with the diffusion calculation, it is worth commenting on the consequences of choosing different initial values of $\alpha$. In Fig.~\ref{Fig_alpha_2} we plot the numerically evaluated $\alpha$ as a function of $\theta$ for different driving times $T$. 
We find that the average value $\langle \alpha\rangle_{\theta}/\alpha_{\theta = \pi/4} \approx 1.24$ for all $T$'s, where $\langle\alpha\rangle_{\theta}$ is obtained by averaging over all $\theta$ values for a particular $T$. Further, we also note that the value of $\alpha$ between $\theta=\pi/4$ and $\theta\approx 0$ differs by just a factor of two. 
Since the entire curve in Fig. 5 shifts vertically as $T$ increases, it follows that changing the initial value of $\alpha$ is tantamount to shifting $T$. This in turn would lead to a horizontal shift of the $T_2$ versus $T$ curves that we obtain in the next section, but it will not affect the shape of these curves. In what follows, we average $\alpha$ over all values of $\theta$ to obtain quantitatively accurate results that can be compared with experimental observations. However, first we will examine qualitative trends in $T_2$ versus $T$, and for this purpose it is sufficient to fix the value of $\alpha$ (here we choose $\alpha=\alpha_{\theta=\pi/4}$) to reduce the computational cost.}


Fourier transforming Eq.~\ref{Eq_I_x0_1}, we have the initial condition
\begin{align}
I(\mathbf{q},0) = \pi^{3/2} \frac{I_0}{\alpha^3} \exp(- ({q_x}^2+ q_y^2 + q_z^2)/4\alpha^2).
\end{align}
We can then calculate the correlation function at finite times using Eq.~\ref{Eq_Iq_Ik}:
\begin{align}
\langle I(\mathbf{q},t)I(\mathbf{k},s) \rangle &= \pi^3\frac{I_0^2}{\alpha^6} e^{-(q^2+k^2)/4\alpha^2 -Dq^2(t+s)}\nonumber \\&- \frac{ \eta}{2} (2\pi)^3  \delta(\mathbf{q}+\mathbf{k}) (e^{-Dq^2(t+s)} - e^{-Dq^2|t-s|}).
\end{align}
{The effect of the net nuclear spin magnetization on a single nuclear spin can be captured using an effective magnetic field description. Consider a nuclear spin at position $x_k$ interacting with all the other nuclear spins via the dipolar interaction, which has the form shown in Eq. (14). The terms that involve the spin operator for the $k$th spin, $I^k$, are
\begin{equation}
H^{k}_{nn} = -\sum_{i\ne k} B_{ik} (I^i_+I^k_- + I^i_-I^k_+) + 2I^k_z \sum_{i\ne k} B_{ik} I^i_z.
\end{equation}
Multiplying by the reduced density matrix for all the nuclear spins other than the $k$th one and performing a partial trace, we arrive at an effective Hamiltonian for the $k$th spin:
\begin{equation}
H^k_{eff} = -\sum_{i\ne k} B_{ik} (\langle I^i_+(t)\rangle I^k_- + \langle I^i_-(t)\rangle I^k_+) + 2I^k_z \sum_{i\ne k} B_{ik} \langle I^i_z(t)\rangle.
\end{equation}
We now assume that only the $z$-component of the nuclear spin bath is nonzero on average so that we retain only the last term above, and we replace the expectation value by the continuous field $I(\mathbf{x},t)$:
\begin{equation}
H^k_{eff} = 2I^k_z \sum_{i\ne k} B_{ik} I(\mathbf{x}_i,t).
\end{equation}
This is essentially the same as performing a semiclassical treatment in which bath correlators are replaced by classical fields (see e.g., Ref. ~\onlinecite{Neder2011}). The coefficients $B_{ik}$ decay like a power law in the distance between the $i$th and $k$th nuclei. When we go to the continuum limit, we are assuming that $I(\mathbf{x},t)$ is approximately constant over distances corresponding to the average inter-nuclear separation, which is on the order of a few Angstroms as estimated above. Since $B_{ik}$ is close to zero for distances much larger than this, it is effectively like a delta-function on the scale on which $I(\mathbf{x},t)$ varies appreciably. Thus, in the continuum limit, we may write
\begin{equation}
H^k_{eff} = \gamma_N I^k_z \int n_k(x) I(x,t),
\end{equation}
where $n(\mathbf{q})$ is the Fourier transform of $n(\mathbf{x})$. This result can be interpreted in terms of an effective magnetic field $B(t)$ seen by the $k$th nuclear spin:}
\begin{align}
B(t) &= \gamma_N\int{d^3\mathbf{x} n(\mathbf{x}) I(\mathbf{x},t)}\nonumber \\
&= \gamma_N\int{\frac{d^3\mathbf{q}}{(2\pi)^3} n(\mathbf{q}) I(-\mathbf{q},t)},
\end{align}
where $n(\mathbf{x})$ is the spatial nuclear density profile of the nucleus, and $n(\mathbf{q})$ is its Fourier transform.  For a given single nucleus located at $(x_0,0,0)$, we assume a Gaussian profile:
\begin{align}
n({\mathbf{x}}) &= \left(\frac{\beta^2}{\pi}\right)^{3/2}\exp(-((x-x_0)^2+y^2+z^2)\beta^2),\\
n(\mathbf{q}) &= \exp(-(q_x^2+q_y^2+q_z^2)/4\beta^2)\exp(-iq_xx_0).
\end{align}
The parameter $\beta$ related to the spatial spread of the nucleus can be adjusted to represent a scenario close to the actual physical case such that $n(\mathbf{x})$ is close to a Dirac-delta function. To obtain physically relevant results, an ensemble averaging over many nuclear sites can be performed by varying the position of the center of the nucleus. For a particular angle $\theta$ the value of the parameter $\alpha$ can be calculated as described previously, and for that chosen angle the radius $x_0$ can be varied within a range. From the definition of $I(\mathbf{k},t)$ the expectation value of the effective magnetic field can be calculated:
\begin{align}
\langle B(t)\rangle = \frac{I_0}{(\alpha)^3}\sqrt{\frac{1}{A^3}}e^{-x_0^2/A},
\label{Eq_mean_B}
\end{align}
where we have defined
$A = {\alpha^{-2}}+{\beta^{-2}}+ 4Dt$.
We are interested in the correlation function for the magnetic field $\langle B(t+s) B(s) \rangle$ which is calculated to be
\begin{align}
\langle B(t+s) B(s) \rangle &=\frac{I_0^2  e^{-x_0^2/F_1}e^{-x_0^2/F_2}}{(\alpha)^6 \sqrt{F_1^3 F_2^3}} \nonumber \\-&\frac{\eta}{2\pi^{3/2}} \left({\frac{1}{F_3^{3/2}}}- {\frac{1}{F_4^{3/2}}}\right), 
\label{Eq_BsptBs_1}
\end{align}
where we have defined the following constants
\begin{align}
F_1 &= {\alpha^{-2}}+{\beta^{-2}}+ 4D (t+2s),\\
\label{Eq_constant1}
F_2 &= {\alpha^{-2}}+{\beta^{-2}},\\
F_3 &= {2}{\beta^{-2}}+4D(t+2s),\\
F_4 &= {2}{\beta^{-2}}+4Dt.
\label{Eq_constant4}
\end{align}
Note that the additional time dependence due to `$s$' in the above equations arises because the nuclear diffusive dynamics start from a non-trivial source term (which comes from the DNP solution) at time `$s$'. This specifically means that we are first driving the system from time $0$ to time $s$, resulting in an initial DNP distribution as discussed earlier. The parameter $\alpha$ is therefore a function of driving time ($\alpha=\alpha(s)$), as also demonstrated in Fig.~\ref{Fig_fit_1}. 
We now define the two point correlation function 
\begin{align}
C(t) = \langle B(s+t) B(s) \rangle - \langle B(s+t) \rangle \langle B(s) \rangle,
\label{Eq_Correl_1}
\end{align}
where the effect of the overall increase of the effective magnetic field $\langle B\rangle$ with  the driving time  has been removed by subtracting the mean value of the effective magnetic field. 
We point out that the Gaussian approximation allows us to analytically calculate the two point correlation function $C(t)$ in Eq.~\ref{Eq_Correl_1}, which otherwise would not be feasible for a more complex form of  $I(\mathbf{x},0)$.
The first spectral density is given by the Fourier transform of $C(t)$  
\begin{align}
C(\omega)=\int\limits_{-\infty}^{+\infty}{dt e^{-i\omega t} C(t)}.
\label{Eq_C_omega}
\end{align}
Since the functional form of $C(t)$ given by Eqs.~\ref{Eq_Correl_1},~\ref{Eq_BsptBs_1},~\ref{Eq_mean_B} does not allow an analytical expression for $C(\omega)$, we will resort to a numerical evaluation of the Fourier transform $C(\omega)$ in our calculations. { In Fig.~\ref{Fig_C_vs_omega} we plot the numerically obtained   $C(\omega)$ for a particular drive time $s$. The inset shows the analytically evaluated  correlation function $C(t)$. Qualitatively we expect the correlation function $C(t)$ to decay as $t\rightarrow \infty$. Numerically we find that $C(\omega)$ is a monotonically decreasing function of $\omega$. Also note that it is the non-trivial variation of $C(\omega)$ with respect to $s$ which is important for determining how $\chi(t)$ and in turn $T_2$ behave, as evaluated in the next section. 
\begin{figure}
	\includegraphics[scale=.6]{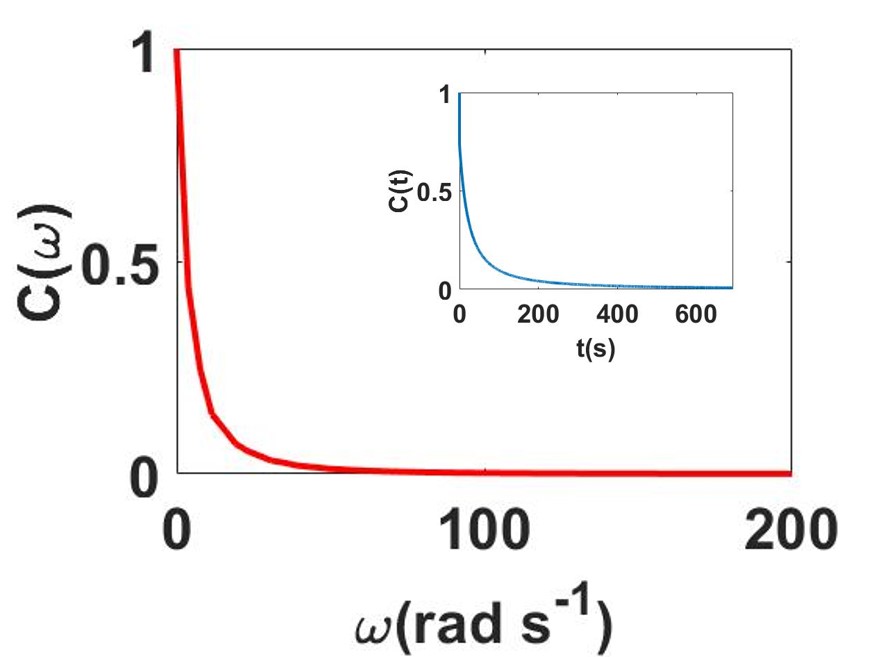}
	\caption{Numerically obtained normalized  $C(\omega)$ for a particular drive time of $s=15$ s. The inset shows the analytically evaluated  correlation function $C(t)$.}
	\label{Fig_C_vs_omega}
\end{figure}
}
\section{Enhancement of nuclear coherence time}
In this section we will use the two point correlation function $C(\omega)$ to evaluate the coherence time $T_2$. Let us consider the quantum state of a single nucleus prepared in an initial state $|\psi\rangle=c_{\uparrow}|\uparrow\rangle+c_{\downarrow}|\downarrow\rangle$ evolving under the stochastic field $B(t)$. The mean field $\langle B(t) \rangle$ and the correlation function of this field were obtained in the previous section (Eq.~\ref{Eq_mean_B} and Eq.~\ref{Eq_BsptBs_1}). The corresponding Hamiltonian is given by $H=\gamma_N B(t) \sigma_z/2$, where $\sigma_z$ is the Pauli matrix. The state at time $t$ will be given by
\begin{align}
|\psi(t)\rangle = e^{-\frac{i}{2}\int\limits_0^t{\gamma_N B(s)}ds}c_\uparrow|\uparrow\rangle + e^{+\frac{i}{2}\int\limits_0^t{\gamma_N B(s)}ds}c_\downarrow|\downarrow\rangle.
\end{align}
The off-diagonal element of the density matrix characterizes the nuclear coherence and can be quantified by the function $W(t)$ as:
\begin{align}
W(t) = \frac{|\langle \rho_{\uparrow\downarrow}(t)\rangle|}{|\langle \rho_{\uparrow\downarrow}(0)\rangle |} = |\langle \exp(-{i}\int\limits_0^t{\gamma_N B(s)}ds)\rangle|.
\end{align}
The function $W(t)$ in the above expression can describe decoherence effects corresponding to a free induction decay, i.e., the nucleus is prepared in a quantum state and allowed to evolve freely under the given Hamiltonian. For more complex pulse sequences (like spin echo, CPMG, PDD etc), a corresponding function $f(t,s)$ can be introduced in the integrand of the above equation to account for multiple pulses~\cite{Cywinski:2008}. The characteristic time of decay of $W(t)$ is denoted as 
$T_2$, defined by $\log (W(T_2))=-1$. Therefore one can typically write the relation $ W(t) \equiv e^{-\chi(t)}$, 
\begin{figure}
\includegraphics[scale=0.45]{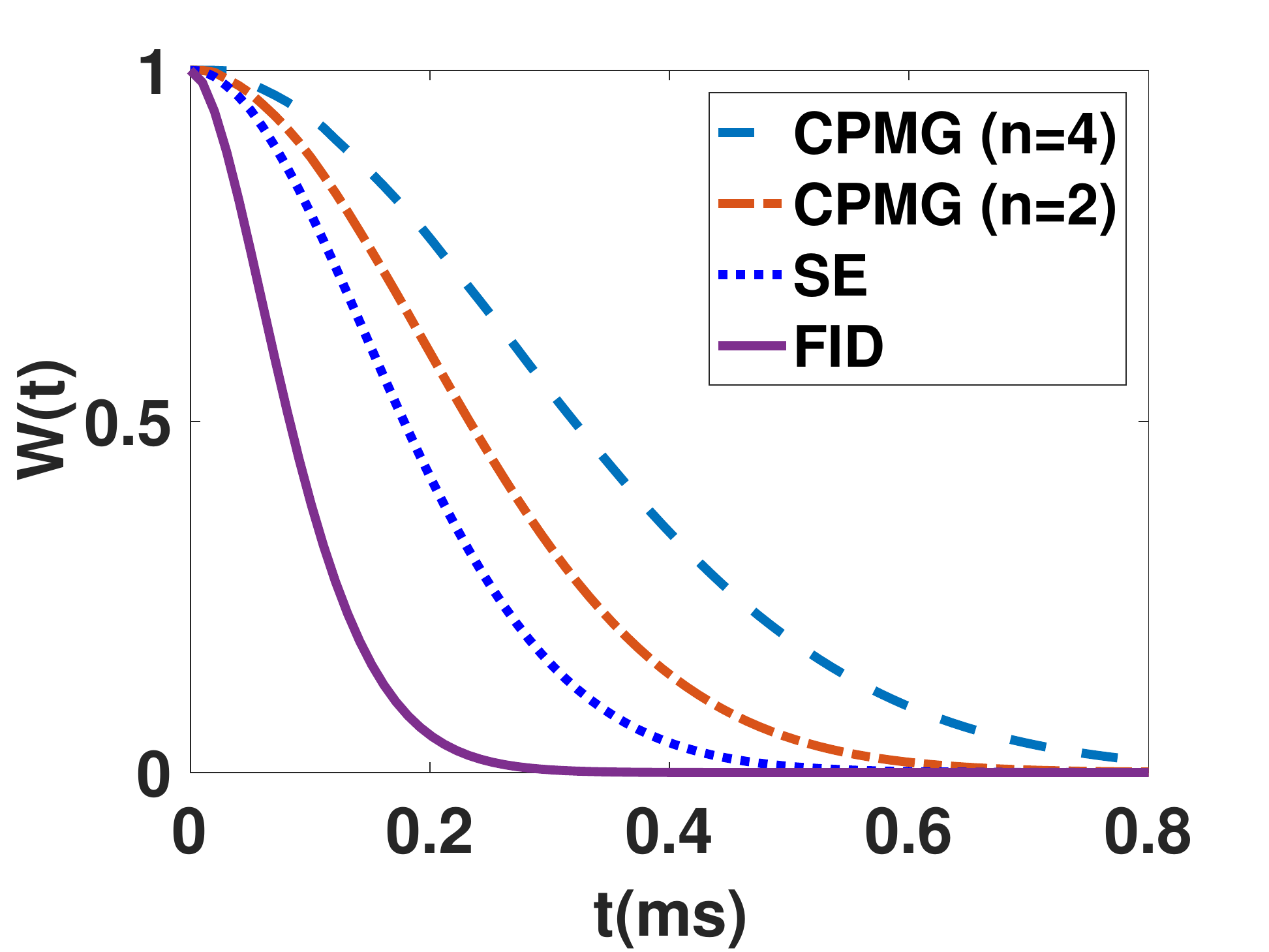}
\caption{Decoherence under various dynamical decoupling pulse sequences (free induction decay, spin echo and CPMG) for a driving time of $T\sim 0.01$s. The other parameters used are $a_1=-0.16632$, $a_2=18.8096$, $x_0=10$nm, $D=25$nm$^2/$s, $\beta^{-1}=1$\AA, $\eta\sim 10^{-20}$(eVs)$^2/$m$^3$, $I_0 = \sqrt{\pi}\hbar/(0.2\mu$m$^3)$. The initial DNP distribution is taken to have $\alpha=\alpha_{\theta=\pi/4}$. }
\label{Fig_W_vs_time}
\end{figure}
 where the quantity $\chi(t)$ can be written as~\cite{Cywinski:2008}
\begin{align}
\chi(t) = \int_0^\infty{\frac{d\omega}{2\pi} C(\omega)\frac{F(\omega t)}{\omega^2}},
\end{align}
where $C(\omega)$ is the first spectral density defined in Sec III Eq.~\ref{Eq_C_omega}. The function $F(z)$ is the filter function, which depends on the type of pulse sequence employed. It encapsulates the effect of the pulse sequence on
decoherence. For example, the filter functions for spin echo (SE) and CPMG pulses are~\cite{Cywinski:2008}
\begin{align}
&F(z) = 8 \sin^4 (z/4) ; \text{ for SE}\\
&F(z) = 8 \sin^4 (z/4n) \sin^2(z/2) / \cos^2 (z/2n) ; \nonumber \\ &\text{ for CMPG even n}.
\end{align}

\begin{figure}
\includegraphics[width=\columnwidth]{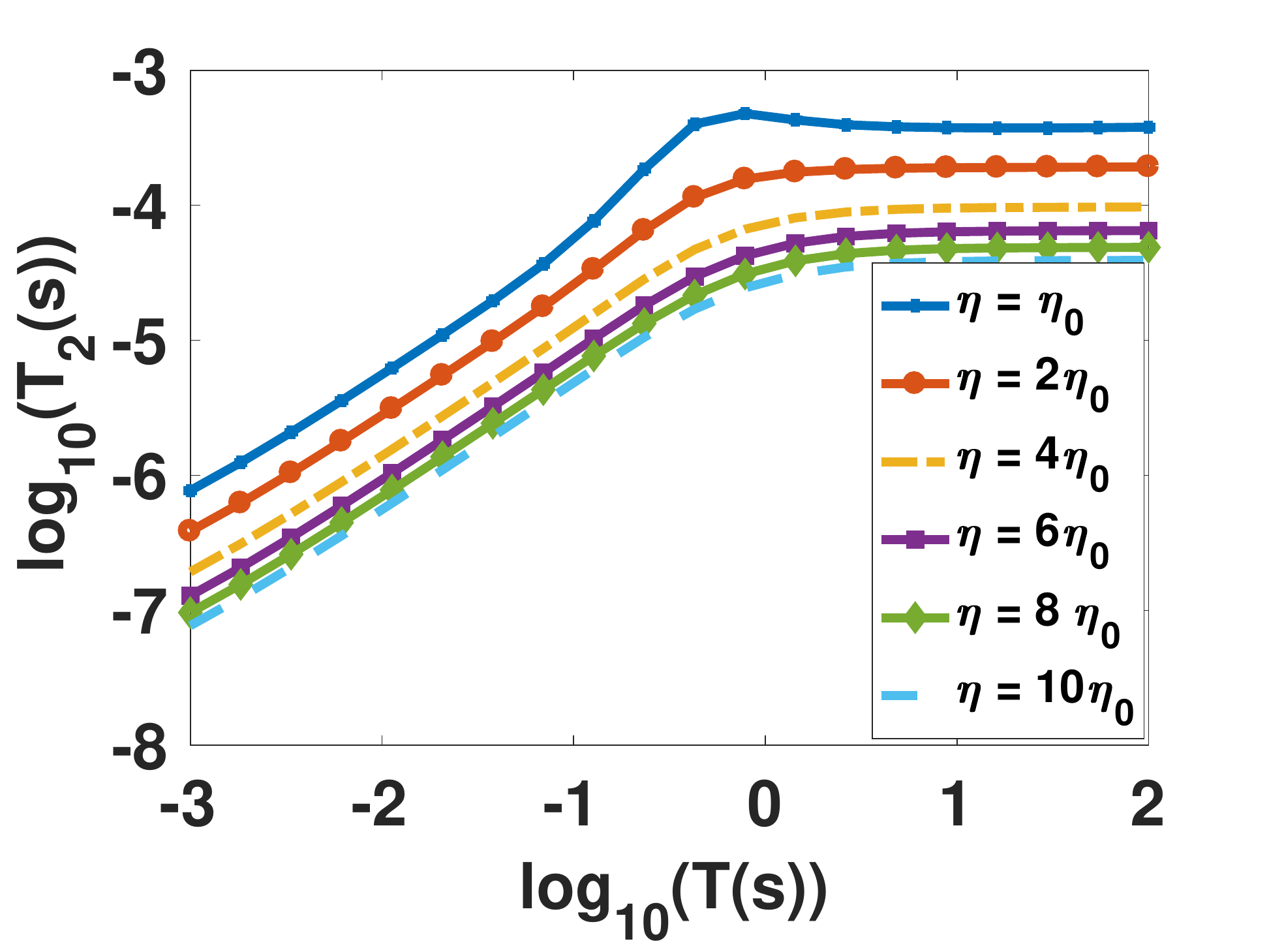}
\caption{Logarithm of the nuclear $T_2$ time as a function of the logarithm of the driving time for various different values of the noise strength $\eta$. A clear enhancement (up to $\sim 3$ orders of magnitude) of the nuclear $T_2$ time with driving time is seen for $-3<\log_{10}(T)<0$. We also note the suppression of $T_2$ time with increasing noise strength. Further the $T_2$ time saturates after the DNP drive time is increased beyond $T\sim 1$s. We use $a_1=-0.16632$, $a_2=18.8096$, $x_0=10$nm, $D=25$nm$^2$/s, $\eta_0\sim 10^{-20}$(eVs)$^2/$m$^3$, $\beta^{-1}=1$\AA, and use the $n=4$ CPMG pulse sequence. The initial DNP distribution is taken to have $\alpha=\alpha_{\theta=\pi/4}$.}
\label{Fig_T2_1}
\end{figure}
\begin{figure}
\includegraphics[width=\columnwidth]{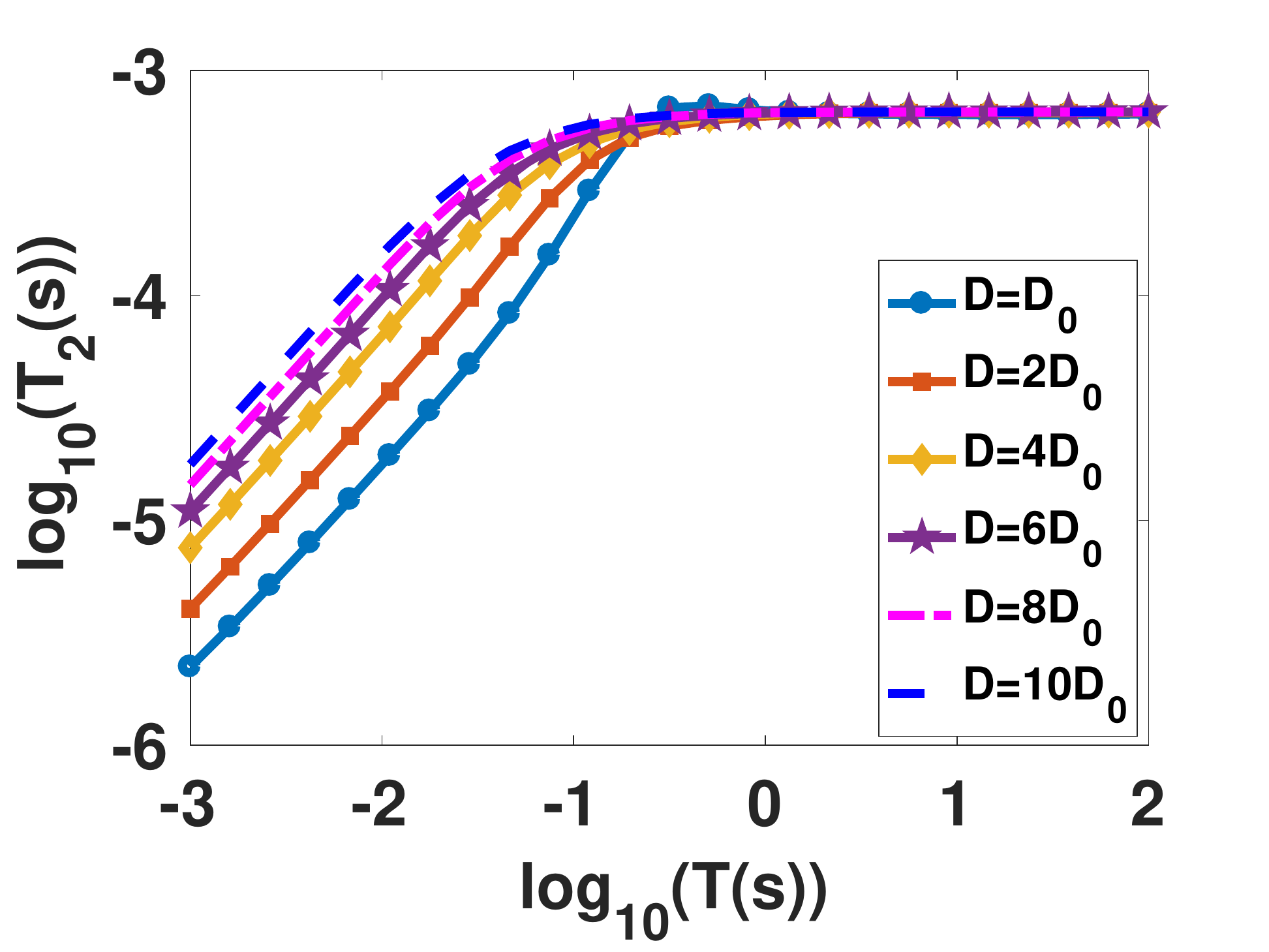}
\caption{Logarithm of the nuclear $T_2$ time as a function of the logarithm of the driving time for various different values of the diffusion constant $D$. For larger $D$ values the coherence time saturates more quickly as a function of driving time. We use $a_1=-0.16632$, $a_2=18.8096$, $x_0=10$nm, $\eta\sim 10^{-20}$(eVs)$^2/$m$^3$, $\beta^{-1}=1$\AA, $D_0=0.5$nm$^2$s$^{-1}$, and use the $n=4$ CPMG pulse sequence. The initial DNP distribution is taken to have $\alpha=\alpha_{\theta=\pi/4}$.}
\label{Fig_T2_2}
\end{figure}
\begin{figure}
\includegraphics[width=\columnwidth]{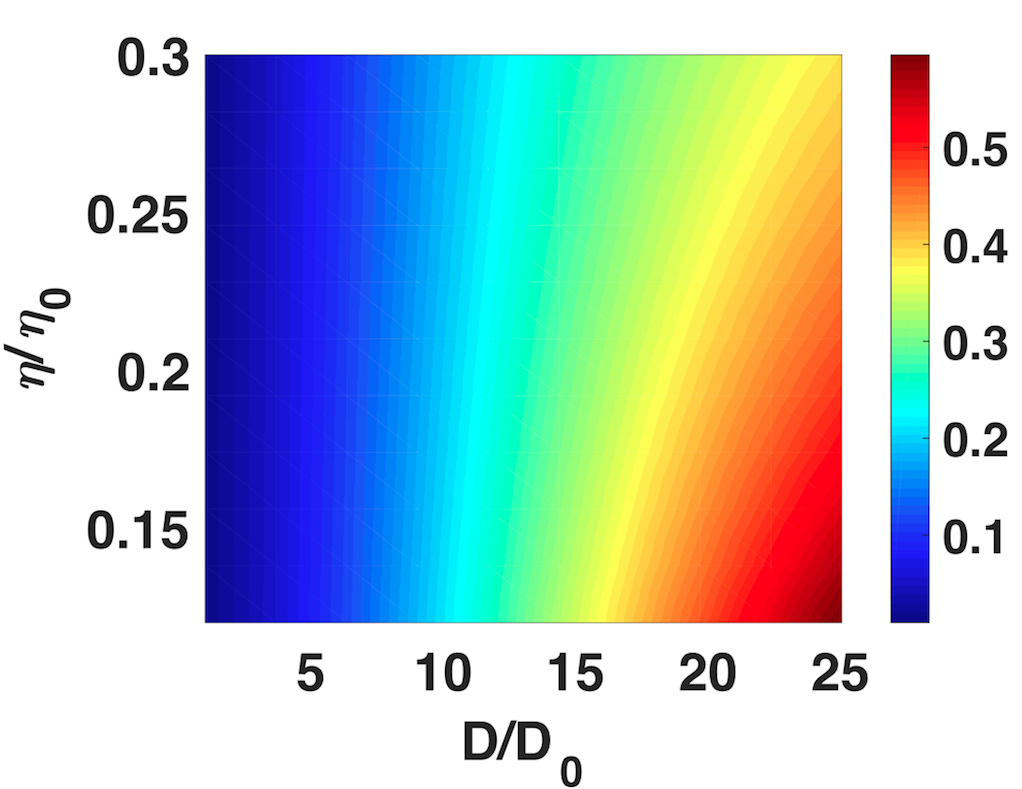}
\caption{Density plot for the nuclear $T_2$ time (in seconds) as a function of the noise parameter and diffusion constant for a constant driving time of $T=10$s. The parameters used were $D_0=25$nm$^2/$s, $\eta_0\sim 10^{-23}$(eVs)$^2$/m$^3$, $a_1=-0.1662$, $a_2=18.4374$, $x_0=10$nm, $\beta^{-1}=1$\AA, and the $n=4$ CPMG pulse sequence was used. The initial DNP distribution is taken to have $\alpha=\alpha_{\theta=\pi/4}$. An increase in noise strength leads to a higher suppression of $T_2$ compared to lowering the diffusion constant $D$. The different order of the $T_2$ obtained here is due to our different choice of parameters compared to Fig.~\ref{Fig_T2_1} and Fig.~\ref{Fig_T2_2}.}
\label{Fig_T2_3}
\end{figure}
\begin{figure}
	\includegraphics[width=\columnwidth]{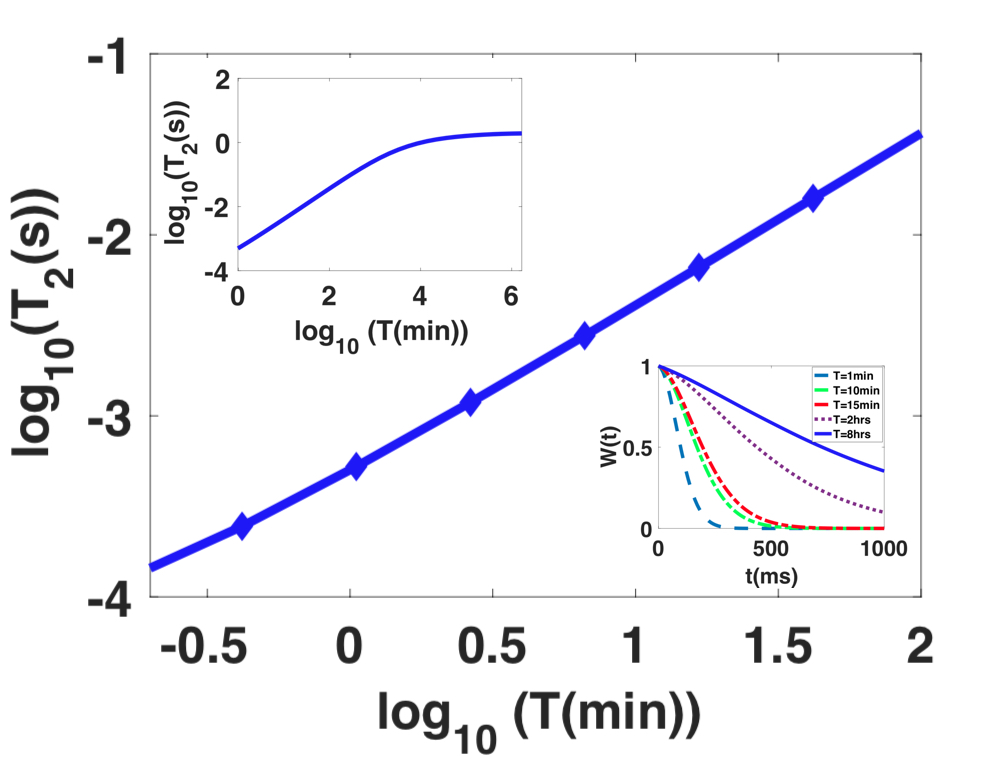}
	\caption{Plot of the logarithm of the nuclear $T_2$ time as a function of logarithm of the DNP polarization time for experimentally relevant parameters for nanodiamond~\cite{Rej:2015} i.e. $D=5$nm$^2/$s, $\omega_{0s}\sim300\mu$eV. Note that for this plot we perform the ensemble averaging to obtain the $T_2$ time over a wide range of radii from the center $0.5$nm$<x_0<10$nm and also over the angle $\theta$. For the DNP Gaussian model for each $\theta$ we have calculated $\alpha$ separately from the exact DNP solution. The top inset shows the saturation of the $T_2$ time for arbitrarily higher DNP driving times and the bottom inset shows the decay of the signal with time for increasing DNP driving times. The other parameters we chose $\beta^{-1}=1$\AA, $\eta\sim 10^{-23}$(eVs)$^2/$m$^3$ ~\cite{Filip:2017}, and the $n=4$ CPMG pulse sequence.}
	\label{Fig_nanodiamond}
\end{figure}

{ In the results, we present below (Fig.~\ref{Fig_W_vs_time},~\ref{Fig_T2_1},~\ref{Fig_T2_2},~\ref{Fig_T2_3}) we will focus on the regime where the nucleus is located at a distance $x_0=10$nm from the electron center. However, for an ensemble measurement, we must average over all possible locations. This is done in Fig.~\ref{Fig_nanodiamond} where we calculate the $T_2$ time for experimentally relevant parameters in nanodiamond.} 
Fig.~\ref{Fig_W_vs_time} shows decoherence under free induction decay and various dynamical decoupling pulse sequences (Hahn's spin echo and CPMG pulses) for a chosen driving time of $T\sim 0.01$s. As one would expect, the dephasing slows down as more dynamical decoupling pulses are applied.

In Fig.~\ref{Fig_T2_1} we show the logarithm of the nuclear $T_2$ as a function of the logarithm of the driving time for various  values of the stochastic noise parameter $\eta$. {For short drive times, we obtain coherence times on the same order of magnitude as prior experimental results for enriched diamond~\cite{Shabanova,Panich}}. We see a clear enhancement  of the nuclear $T_2$ time with driving time for $-3<\log_{10}(T)<0$. The $T_2$ time is suppressed with increasing noise parameter $\eta$, however we point out that it is essential that $\eta \neq 0$ in order to obtain physically acceptable results for $T_2$. 
Further we also note that the $T_2$ time saturates after the driving time is increased beyond a certain time $T_{sat}$. The exact value of the obtained $T_2$ time and the saturation time depends on our choice of parameters. Fig.~\ref{Fig_T2_2} shows the logarithm of the nuclear $T_2$ as a function of the logarithm of the driving time for various values of the diffusion constant $D$. We note that for a higher diffusion constant, the $T_2$ coherence time saturates more quickly as a function of driving time. This behavior is expected because a higher diffusion rate should increase the temporal spread of the driving induced polarization (see Eq.~\ref{Eq_Diff_1}), however $\eta$ sets the scale for an upper limit on $T_2$. Fig.~\ref{Fig_T2_3} shows the density plot for the nuclear $T_2$ time as a function of the noise strength and diffusion constant for a constant driving time $T$. An increase in noise strength leads to a higher suppression of $T_2$ compared to lowering the diffusion constant $D$. The different order of the $T_2$ obtained in Fig.~\ref{Fig_T2_3} is due to our different choice of parameters compared to Fig.~\ref{Fig_T2_1} and Fig.~\ref{Fig_T2_2}.

The two processes that play a central role in our analysis are DNP induced by driving electron-nuclear flip-flops and diffusion caused by nuclear dipole-dipole interactions. For finite magnetic fields the dipolar interaction between two nuclei can be effectively written as a sum of Overhauser and flip-flop terms $H_{dip} \approx t' (I_{+i}I_{-j} + I_{+j}I_{-i}- 2I_{zj}I_{zi})$, where $t'$ is the energy scale of the interaction, and $i,j$ represent nuclei indices. When the nuclear bath is completely unpolarized ($m=0$), the distribution of the nuclear spins in the configuration space has the maximum entropy, while for a fully polarized nuclear bath ($m=1$) the configurational entropy is zero as there is only one way to arrange the spins. Therefore when we consider processes in which the total spin is conserved, such as a pair of spin-flips which causes fluctuations in the magnetic field (noise), an unpolarized bath is expected to result in maximal noise and therefore the lowest coherence time $T_2$, while a fully polarized bath should result in minimal noise and a maximal (ideally infinite if other decoherence processes are ignored) $T_2$ time. For intermediate bath polarizations, the phase space for flip-flops is reduced and is sharply peaked around $m=0$. Therefore physically we also expect that longer driving times, which causes higher average bath polarization, should result in an enhanced $T_2$ coherence time. 
Further we note that all the curves in Fig.~\ref{Fig_T2_2} saturate to the same value of $T_2$ indicating the fact that the nuclear polarization itself saturates to the same distribution regardless of the diffusion constant. The magnitude of the diffusion constant only affects how quickly the DNP reaches its saturation value. The $T_2$ time for a nucleus at $x=10$ nm only improves once the DNP has propagated till $x_0$ from the electron defect center, after which the $T_2$ time remains a constant as the DNP wave continues to propagate outwards, and the DNP becomes uniform over the sample. { The fact that all the curves in Fig.~\ref{Fig_T2_1} also saturate near the same value of $T$ is again a consequence of the fact that the DNP saturation point only depends on $T$ and not on $\eta$. However, the different curves in Fig.~\ref{Fig_T2_1} do not all saturate to the same value of $T_2$ because the dephasing caused by $\eta$ diminishes the gains in $T_2$ afforded by the DNP.}

In Fig.~\ref{Fig_W_vs_time}-~\ref{Fig_T2_3} we demonstrated the generic behavior of the $T_2$ time as a function of the driving time as obtained within our theoretical model. In actual hyperpolarization experiments, the driving time can be increased up to a few minutes or even hours. In recent experiments~\cite{Rej:2015} performed on nanodiamond, an increase of the relaxation time of $^{13}$C spins up to 3 orders of magnitude has been observed by dynamically hyperpolarizing the nuclear bath via microwave driving. In Fig.~\ref{Fig_nanodiamond} we plot the $T_2$ coherence time using experimentally relevant parameters for nanodiamond and note that the nuclear coherence time increases up to 3 orders of magnitude ($\sim 1$ms-$1$s) as the driving time is increased. Note that for this plot we perform the ensemble averaging to obtain the $T_2$ time over a wide range of radii from the center $0.5$nm$<x_0<10$nm for a particular $\theta$ and also average over the angle $\theta$. Since the DNP solution is anisotropic and has a $C_4$ symmetry, for the Gaussian model for each $\theta$ we have calculated $\alpha$ separately from the exact DNP solution. {Our theoretical results suggest that nuclear hyperpolarization via microwave driving not only enhances relaxation times but also nuclear spin coherence times by several orders of magnitude, as also suggested by preliminary experimental results on nanodiamond $^{13}$C spins~\cite{Expt,aps}}. 

\section{Conclusion}
In this work, we calculated the nuclear spin coherence time for an ensemble of dipolar-coupled nuclear spins in the vicinity of a driven defect center in a solid. We showed that when electron-nuclear spin-flip transitions are driven with microwave fields, nonsecular terms in the electron-nuclear hyperfine interaction can generate a large dynamic nuclear polarization. { We then analytically calculated the spatial distribution of this nuclear polarization as a function of the driving time and used this as a starting point to study the subsequent diffusion and fluctuations of the polarization. To study the evolution nuclear $T_2$ time in an analytically tractable manner, we approximate the spatial distribution of the nuclear polarization around the defect center by a Gaussian distribution.} 
Using these results, we then obtained the coherence times of nuclear spins far from the defect center as a function of the driving time, fluctuation strength, and speed of diffusion. We found that the coherence generically increases by several orders of magnitude as the driving time is increased up until a saturation point that depends on the strength of the dipolar interaction. In the case of $^{13}$C nuclear spins, this translates to a nearly three orders of magnitude coherence time increase, a result that parallels a similar enhancement in relaxation times seen in recent experiments\cite{Rej:2015}. {Our theoretical model and results will be therefore useful for current and upcoming experiments on enhancing the coherence time via DNP. Preliminary experimental results on nanodiamond $^{13}$C spins also suggest that the nuclear spin coherence time is in fact enhanced by the hyperpolarization~\cite{Expt,aps}.} 

\textit{Acknowledgment:} The work of E.B. and S. E. E. was supported by NSF (Grant No. 1741656).

\pagebreak

\end{document}